\documentclass[aps,prl,twocolumn,showpacs,amsmath,amsmath,amssymb,amsfonts]{revtex4-1}
\usepackage{graphicx}
\usepackage{dcolumn}
\usepackage{bm}
\usepackage{color}
\usepackage{multirow}

\begin{document}
\author{Igor Lesanovsky$^{1, 2, 3}$, Beatriz Olmos$^{1,2}$, William Guerin$^{3}$ and Robin Kaiser$^{3}$}
\affiliation{$^{1}$School of Physics and Astronomy, University of Nottingham, Nottingham, NG7 2RD, UK}
\affiliation{$^{2}$Centre for the Mathematics and Theoretical Physics of Quantum Non-equilibrium Systems,
University of Nottingham, Nottingham NG7 2RD, UK}
\affiliation{$^{3}$Universit\'e C\^ote d'Azur, CNRS, INPHYNI, France}
\title{Dressed dense atomic gases}
\date{\today}
\begin{abstract}
In dense atomic gases the interaction between transition dipoles and photons leads to the formation of many-body states with collective dissipation and long-ranged forces. Despite decades of research, a full understanding of this paradigmatic many-body problem is still lacking. Here, we put forward and explore a scenario in which a dense atomic gas is weakly excited by an off-resonant laser field. We develop the theory for describing such dressed many-body ensembles and show that collective excitations are responsible for the emergence of many-body interactions, i.e. effective potentials that cannot be represented as a sum of binary terms. We illustrate how interaction effects may be probed through microwave spectroscopy via the analysis of time-dependent line-shifts, and show that these signals are sensitive to the phase pattern of the dressing laser. Our study offers a new perspective on dense atomic ensembles interacting with light and promotes this platform as a setting for the exploration of rich non-equilibrium many-body physics.
\end{abstract}


\maketitle
\textit{Introduction ---} There is currently substantial interest in probing and understanding the physics of dense atomic ensembles, as they represent a paradigmatic many-body system whose properties are governed solely by the fundamental interaction of photons with matter. These systems feature strong dipole-dipole interactions, which have been at the center of intense research in atomic physics for many decades. Contact and long range two-body dipole-dipole interactions have been investigated in the context of cold molecular physics \cite{Weiner1999}, to identify novel states based on two particle entanglement \cite{Hettich2002} or for their impact for high precision measurement with cold neutral atoms \cite{Chang2004, Ludlow2015}.
A number of experimental results have been obtained via direct laser spectroscopy. Those include the observation of the collective Lamb-shift and the Lorentz-Lorenz shift \cite{Friedberg1973,Maki1991,Sautenkov1996,scully2009,keaveney2012,Peyrot2018} as well as the emergence of sub-\cite{bienaime2012,guerin2016} and superradiance~\cite{scully2009,araujo2016,roof2016,Solano2017}. These results have confirmed theoretical predictions made as early as in the 1950's \cite{dicke1954}, but at the same time led to further questions. A particular example for this is the current difficulty in reconciling observed level shifts with established textbook knowledge \cite{jenkins2016,bromley2016}. This clearly signals further need for investigating the collective behavior of these many-body systems and for developing new ways of probing and exploring their collective dynamics.
\begin{figure}[t!]
  \includegraphics[width=0.9\columnwidth]{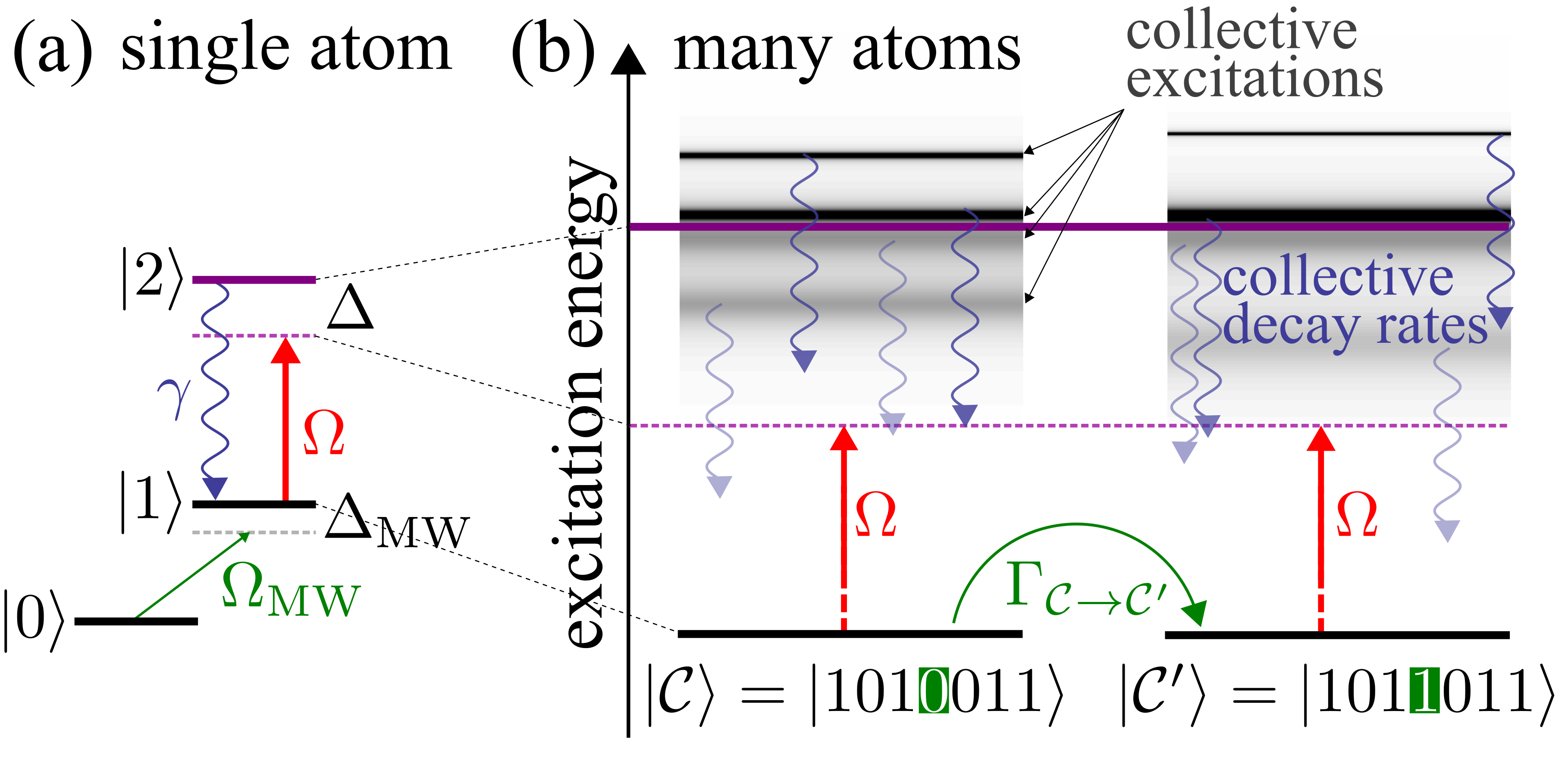}
  \caption{\textbf{Dressed atomic gas.} (a) Sketch of the single-atom level structure. The off-resonant dressing laser (Rabi frequency $\Omega$, detuning $\Delta$) weakly drives the transition $\left|1\right>\rightarrow\left|2\right>$ that has a decay rate $\gamma$.
  The resulting level shifts can be probed via microwave (MW) spectroscopy on the $\left|0\right>\rightarrow\left|1\right>$ transition. The Rabi frequency and detuning of the MW field are $\Omega_\mathrm{MW}$ and $\Delta_\mathrm{MW}$, respectively. (b) Atoms exchange (virtual) photons on the transition $\left|1\right>\rightarrow\left|2\right>$, which lead to the formation of delocalised many-body states (collective excitations) with collective decay rates. Sketched are the collective energy levels for two different atomic configurations, $\left|\mathcal{C}\right>$ and $\left|\mathcal{C}'\right>$. The many-body level structure depends on the number and spatial arrangement of atoms in the state $\left|1\right>$, and the energies of the collective excitation states are shifted and broadened with respect to the single atom state. The MW field effectuates transitions between the states $\left|\mathcal{C}\right>$ and $\left|\mathcal{C}'\right>$ at a rate $\Gamma_{\mathcal{C}\rightarrow\mathcal{C}'}$ which permits a spectroscopic analysis of the dressed state manifold.}
\label{fig:1}
\end{figure}

In this work we develop a theory for dressed dense atomic gases, a scenario in which excited atomic states are weakly coupled to the atomic ground state via an off-resonant laser. Laser-dressing has become popular in recent years, in particular in the context of Rydberg atoms \cite{jau2016,zeiher2016}, where it is used to tailor interaction potentials for the purpose of simulating exotic types of matter \cite{vanbijnen2015,glaetzle2015} or the generation of entangled many-body states \cite{bouchoule2002,gil2014,arias2018} for quantum enhanced measurements. We show here that dressing of dense atomic gases may produce strong effective interactions already at the level of second order in the strength of the dressing laser (in contrast to Rydberg gases where this is a fourth-order effect). The resulting effective interatomic potential is in general of many-body type, i.e. it cannot be decomposed as the sum of binary interaction terms. Moreover, dissipation and coherent interaction are inextricably interlinked, which necessitates the application of a perturbative treatment within an open system formulation. We discuss how the arising collective properties can be probed via microwave (MW) spectroscopy, through the shift and time-evolution of the absorption line. Beyond shedding light on fundamental aspects of the interaction between matter and photons, our results are of relevance in the context of technological devices based on dense atomic ensembles, such as lattice clocks \cite{loic2019}.

\textit{Many-body model ---} The model we are considering here consists of $N$ three-level atoms located at positions $\mathbf{r}_\alpha$. As shown in Fig. \ref{fig:1}a, the levels $\left|1\right>$ and $\left|2\right>$ are weakly coupled with an off-resonant dressing laser with detuning $\Delta$ and single atom Rabi frequencies $\Omega_\alpha$ ($|\Omega_\alpha|\ll|\Delta|$), which may be complex and also different from atom to atom. The resulting (light-)shift and decay rate of atoms from state $\left|2\right>$ into state $\left|1\right>$ can then be probed by a MW field on the $\left|0\right>\rightarrow\left|1\right>$ transition.

For the time being we take the MW field to be absent and focus on the dynamics that takes place between the levels $\left|1\right>$ and $\left|2\right>$. This is governed by the dressing laser and its interplay with the inter-particle interactions, which are obtained by integrating out the free photon modes \cite{lehmberg1970,bienaime2011,bienaime2013}. The corresponding equation of motion is a Markovian master equation of the form $\partial_t \rho= \mathcal{L} \rho$. Here $\rho$ is the many-body density matrix of the atomic system and $\mathcal{L}$ is the master operator which we decompose as $\mathcal{L}=\mathcal{L}_0+\mathcal{L}_1$. The two terms correspond to a "fast" and a "slow" dynamics. The slow dynamics is given by the coupling to the dressing laser, represented by the operator
\begin{eqnarray}
  \mathcal{L}_1\bullet=-i\sum_\alpha^N \left[\Omega_\alpha^* b_\alpha^+ + \Omega_\alpha b_\alpha,\bullet\right],\label{eq:L1}
\end{eqnarray}
while the fast dynamics is given by
\begin{eqnarray}
  \mathcal{L}_0\bullet&=&-i\sum_k^N \Delta \left[b_\alpha^+ b_\alpha, \cdot\right] -i\sum_{\alpha\neq \beta}^NV_{\alpha \beta}\left[b^+_\alpha b_\beta,\bullet\right]\nonumber\\
  &&+\sum_{\alpha \beta}^N G_{\alpha \beta}\left(b_\alpha \bullet b_\beta^+ - \frac{1}{2}\left\{b_\beta^+ b_\alpha,\bullet\right\}\right),\label{eq:L0}
\end{eqnarray}
which depends on the interaction and dissipation matrices
\begin{eqnarray}
  V_{\alpha \beta}&=&\frac{3\gamma}{4}\left\{-\left[1-(\hat{d}\cdot\hat{r}_{\alpha\beta})^2\right]\frac{\cos\kappa_{\alpha \beta}}{\kappa_{\alpha \beta}}\right.\label{eq:V_matrix}\\
  &&\left.+\left[1-3(\hat{d}\cdot\hat{r}_{\alpha\beta})^2\right]\left(\frac{\sin\kappa_{\alpha \beta}}{\kappa_{\alpha \beta}^2}+\frac{\cos\kappa_{\alpha \beta}}{\kappa_{\alpha \beta}^3}\right)\right\}\nonumber\\
  G_{\alpha \beta}&=&\frac{3\gamma}{2}\left\{\left[1-(\hat{d}\cdot\hat{r}_{\alpha\beta})^2\right]\frac{\sin\kappa_{\alpha \beta}}{\kappa_{\alpha \beta}}\right.\label{eq:G_matrix}\\
  &&\left.+\left[1-3(\hat{d}\cdot\hat{r}_{\alpha\beta})^2\right]\left(\frac{\cos\kappa_{\alpha \beta}}{\kappa_{\alpha \beta}^2}-\frac{\sin\kappa_{\alpha \beta}}{\kappa_{\alpha \beta}^3}\right)\right\}\nonumber.
\end{eqnarray}
The matrices $V_{\alpha \beta}$ and $G_{\alpha \beta}$ are functions of the reduced length $\kappa_{\alpha\beta}=kr_{\alpha\beta}$, product of the separation between the atoms $r_{\alpha \beta}=|\mathbf{r}_\alpha-\mathbf{r}_\beta|$ and the wave number $k$, corresponding to the atomic transition $\left|1\right>\rightarrow\left|2\right>$. They furthermore depend on the decay rate $\gamma$ of this transition and the relative direction of its dipole moment $\hat{d}$ and the interatomic separations $\hat{r}_{\alpha \beta}=(\mathbf{r}_\alpha-\mathbf{r}_ \beta)/r_{\alpha \beta}$.

In formulating the master equation we have used the operators $b^+_\alpha=\left|2\right>_\alpha\!\!\left<1\right|$ that describe the excitation of the atom at position $\mathbf{r}_\alpha$ from state $\left|1\right>$ to $\left|2\right>$. In the following we will assume that these operators obey bosonic commutation relations, i.e. $\left[b_\alpha,b_\beta^+\right]=\delta_{\alpha \beta}$, which is justified when the number of excited atoms is small compared to the total number of atoms $N$. Within the relevant leading order of the subsequent perturbative treatment, this approximation is actually exact.

\textit{Dressed many-body states and effective equation of motion ---}
Each (classical) atomic configuration of the form
\begin{eqnarray}
\left|\mathcal{C}\right>=\otimes_\alpha\left|\xi_\alpha\right> \label{eq:configuration}
\end{eqnarray}
with $\xi_\alpha=0,1$ evolves independently and is weakly coupled (via $\mathcal{L}_1$) to a set of collective many-body states whose coherent and dissipative dynamics is governed by the master operator $\mathcal{L}_0$. Since $\mathcal{L}_0$ couples only the states $\left|1\right>$ and $\left|2\right>$, the form of the excited collective states is dictated by the number as well as the arrangement of atoms in state $\left|1\right>$ of a given configuration $\left|\mathcal{C}\right>$. This is sketched for two configurations in Fig. \ref{fig:1}b. The dressing makes each configuration acquire a collective energy shift and decay rate but does not couple different configurations of the form (\ref{eq:configuration}).

In order to understand the dynamics of the dressed atomic ensemble we seek an effective equation of motion for the density matrix $\mu=\sum_{\mathcal{C}\mathcal{C}'}\mu_{\mathcal{C}\mathcal{C}'}\left|\mathcal{C}\right>\!\left<\mathcal{C}'\right|$. In the following we outline the main steps of the calculation, leading to the main result given by Eq. (\ref{eq:effective_evolution}). We begin with the general formula for second order perturbation theory \cite{breuer2002}:
\begin{eqnarray}
  \partial_t \mu &=& \mathcal{L}_\mathrm{eff}\mu=\mathcal{P}\!\! \int_0^\infty dt\, \mathcal{L}_1 e^{\mathcal{L}_0 t}\mathcal{L}_1\mu\\
  &=&\! -\mathcal{P}\!\! \int_0^\infty \!\!\!\!\!dt \sum_{\gamma\alpha}\!\!\left[\Omega_\gamma^* b_\gamma^+ + \Omega_\gamma b_\gamma,e^{\mathcal{L}_0 t}\!\!\left(\left[\Omega_\alpha^* b_\alpha^+ + \Omega_\alpha b_\alpha,\mu\right]\right)\right].\nonumber
\end{eqnarray}
Here $\mathcal{P}=\lim_{t\rightarrow\infty}e^{\mathcal{L}_0 t}$ is the projector on the stationary subspace of $\mathcal{L}_0$, which is spanned by the configurations (\ref{eq:configuration}). To carry out the calculation explicitly, we focus on the evolution of the basis vectors $\left|\mathcal{C}\right>\!\left<\mathcal{C}'\right|$. We exploit that $b_\alpha \left|\mathcal{C}\right>\!\left<\mathcal{C}'\right|= \left|\mathcal{C}\right>\!\left<\mathcal{C}'\right|b^+_\alpha=0$ and furthermore
\begin{eqnarray*}
  \partial_t \left(b^+_\alpha\left|\mathcal{C}\right>\!\left<\mathcal{C}'\right|\right) = \mathcal{L}_0 \left(b^+_\alpha\left|\mathcal{C}\right>\!\left<\mathcal{C}'\right|\right)=\sum_{\beta} M^\mathcal{C}_{\alpha\beta} \left(b^+_\beta\left|\mathcal{C}\right>\!\left<\mathcal{C}'\right|\right),
\end{eqnarray*}
with the symmetric matrix $\mathbf{M}^\mathcal{C}$, whose components are
\begin{eqnarray}
  M^\mathcal{C}_{\alpha\beta}=-i\Delta\delta_{\alpha\beta} -i V_{\alpha\beta} -\frac{1}{2} G_{\alpha\beta}\label{eq:M_matrix}
\end{eqnarray}
and involve the coefficients (\ref{eq:V_matrix}) and (\ref{eq:G_matrix}). Here the superscript $\mathcal{C}$ emphasizes that the $\mathbf{M}^\mathcal{C}$ depends on the structure of configuration $\left|\mathcal{C}\right>$, i.e. on the number and arrangement of atoms in state $\left|1\right>$. Integrating the equation of motion for  the operators $b^+_\alpha\left|\mathcal{C}\right>\!\left<\mathcal{C}'\right|$ yields
\begin{eqnarray}
  \mathcal{P} \int_0^\infty dt\, b_\beta e^{\mathcal{L}_0 t}\left(b^+_\alpha\left|\mathcal{C}\right>\!\left<\mathcal{C}'\right|\right)=-\Lambda^\mathcal{C}_{\alpha\beta} \left|\mathcal{C}\right>\!\left<\mathcal{C}'\right|,\nonumber
\end{eqnarray}
with the matrix $\mathbf{\Lambda}^\mathcal{C}=-\int_0^\infty dt\, e^{t\mathbf{M}^\mathcal{C}}=\left(\mathbf{M}^\mathcal{C}\right)^{-1}$.
Here we have exploited that $\mathcal{P} b_\alpha b^+_\beta
\left|\mathcal{C}\right>\!\left<\mathcal{C}'\right|=\delta_{\alpha\beta}\left|\mathcal{C}\right>\!\left<\mathcal{C}'\right|$.
The computation of the other terms is analogous and involves the expression
$\mathcal{P} b^+_\alpha \left|\mathcal{C}\right>\!\left<\mathcal{C}'\right| b_\beta=\Theta^{\mathcal{C}\mathcal{C}'}_{\alpha\beta}\left|\mathcal{C}\right>\!\left<\mathcal{C}'\right|$ where
\begin{eqnarray*}
  \Theta^{\mathcal{C}\mathcal{C}'}_{\alpha\beta}=\left[\int^\infty_0 ds\, e^{s \mathbf{M}^\mathcal{C}} \mathbf{G}^{\mathcal{C}\mathcal{C}'} e^{s{\mathbf{M}^{\mathcal{C}'}}^*} \right]_{\alpha\beta}.
\end{eqnarray*}
The coefficients of the matrix $G_{\alpha\beta}^{\mathcal{C}\mathcal{C}'}$ are given by Eq. (\ref{eq:G_matrix}), and the indices $\alpha$ and $\beta$ run through the atoms in state $\left|1\right>$ of the configurations $\left|\mathcal{C}\right>$ and $\left|\mathcal{C}'\right>$, respectively. Note that the integral $\Theta^{\mathcal{C}\mathcal{C}'}_{\alpha\beta}$ can be solved via matrix inversion and thus no explicit integration needs to be performed in numerical simulations.

Putting all terms together yields the effective equation of motion for the density matrix basis states:
\begin{eqnarray}
  \partial_t \left|\mathcal{C}\right>\!\left<\mathcal{C}'\right| =  \mathcal{L}_\mathrm{eff} \left|\mathcal{C}\right>\!\left<\mathcal{C}'\right| = -i\Delta_{\mathcal{C}\mathcal{C}'} \left|\mathcal{C}\right>\!\left<\mathcal{C}'\right|.\label{eq:effective_evolution}
  \end{eqnarray}
This equation shows that each element $\left|\mathcal{C}\right>\!\left<\mathcal{C}'\right|$ evolves according to the (complex) energy difference
\begin{eqnarray}
  -i\Delta_{\mathcal{C}\mathcal{C}'}&=&\vec{\Omega}^*_\mathcal{C}\cdot \mathbf{\Lambda}^\mathcal{C}\cdot\vec{\Omega}_\mathcal{C} + {\vec{\Omega}_{\mathcal{C}'}}\cdot{\mathbf{\Lambda}^{\mathcal{C}'}}^* \cdot\vec{\Omega}^*_{\mathcal{C}'}\\
  &&-\vec{\Omega}^*_\mathcal{C} \cdot \mathbf{\Lambda}^\mathcal{C} \cdot \mathbf{\Theta}^{\mathcal{C}\mathcal{C}'}\cdot{\vec{\Omega}_{\mathcal{C}'}}-\vec{\Omega}^*_\mathcal{C} \cdot \mathbf{\Theta}^{\mathcal{C}\mathcal{C}'}  \cdot {\mathbf{\Lambda}^{\mathcal{C}'}}^*  \cdot{\vec{\Omega}_{\mathcal{C}'}},\nonumber
\end{eqnarray}
where the vectors $\vec{\Omega}_\mathcal{C}$ contain the dressing laser Rabi frequencies of all atoms in the state $\left|1\right>$ contained in  $\left|\mathcal{C}\right>$. The real and imaginary part of  $\Delta_{\mathcal{C}\mathcal{C}'}$ are the energy difference and dephasing rate of the configurations $\left|\mathcal{C}\right>$ relative to $\left|\mathcal{C}'\right>$ as we will further discuss below. Note that $\Delta_{\mathcal{C}\mathcal{C}}=0$, i.e. the diagonal elements of the density matrix $\mu$ do not evolve in time.

\begin{figure}[t!]
  \includegraphics[width=0.7\columnwidth]{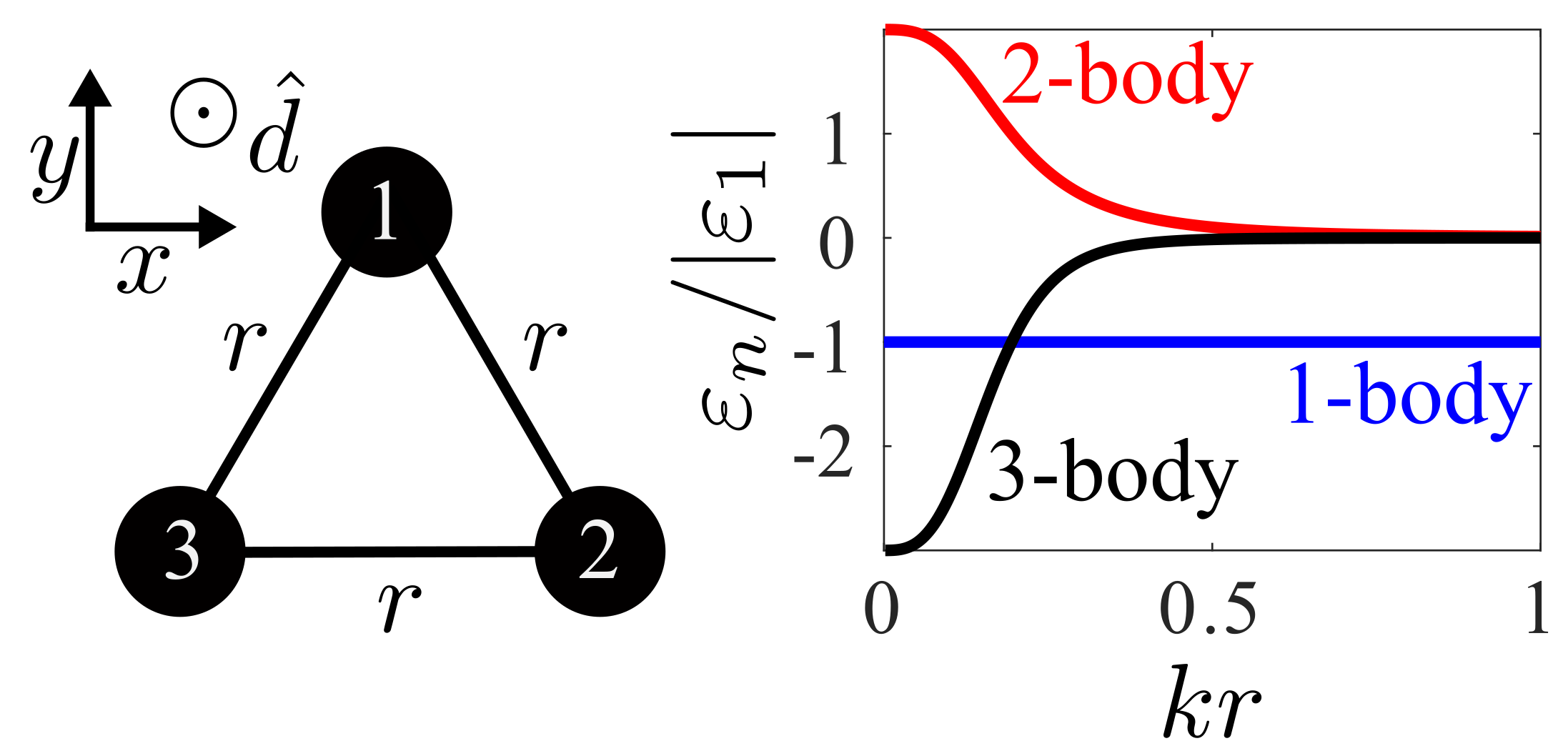}
  \caption{\textbf{Few-body interactions.} Three atoms on an equilateral triangle adressed by a laser with uniform Rabi frequency, $\Omega=5\gamma$, and detuning $\Delta=100\gamma$. The atoms are positioned in the $xy$-plane and the atomic dipole moment is pointing into the $z$-direction. The right panel shows the single, two- and three-body contribution, $\varepsilon_n$, to the (interaction) energy.}
\label{fig:2}
\end{figure}
\textit{Few atom system ---} Let us now consider a system where three atoms are positioned on an equilateral triangle, as shown in Fig. \ref{fig:2}. The atoms are positioned in the $xy$-plane and the dipole moment of the $\left|1\right>\rightarrow\left|2\right>$ transition points into the $z$-direction, i.e. $\hat{d}=\mathbf{e}_z$. For the sake of simplicity we assume that the dressing laser Rabi frequency is uniform and real, i.e. we neglect the spatial dependence of the laser phase. The question now is whether the effective interacting energy is the sum of binary interactions or whether there are three-body interactions involved. The single-body energy is given by
\begin{eqnarray}
  \varepsilon_1=-\mathrm{Re}\left[\Delta_{\left|000\right>\left|100\right>}\right]=-\frac{4\Delta\Omega^2}{\gamma^2+4\Delta^2},
\end{eqnarray}
which is simply the light shift. Analogously, we obtain the two-atom interaction potential
\begin{eqnarray}
  \varepsilon_2&=&-\mathrm{Re}\left[\Delta_{\left|100\right>\left|110\right>}\right]-\varepsilon_1\nonumber\\
  &=&\frac{8\Omega^2}{\gamma^2+4\Delta^2}\frac{(V_{12}+\Delta)(\gamma G_{12}+4\Delta V_{12})}{(\gamma+G_{12})^2+4(\Delta+V_{12})^2}.
\end{eqnarray}
As one can observe in Fig. \ref{fig:2}, for very small interparticle distances, $kr\ll 1$, $\varepsilon_2$ shows a characteristic flat-top shape, and for large $kr$ one finds $\varepsilon_2\rightarrow \frac{2\Omega^2}{\Delta^2} V_{12}$, i.e. the effective interaction potential becomes proportional to Eq. (\ref{eq:V_matrix}).

\begin{figure*}[t!!]
  \includegraphics[width=1.6\columnwidth]{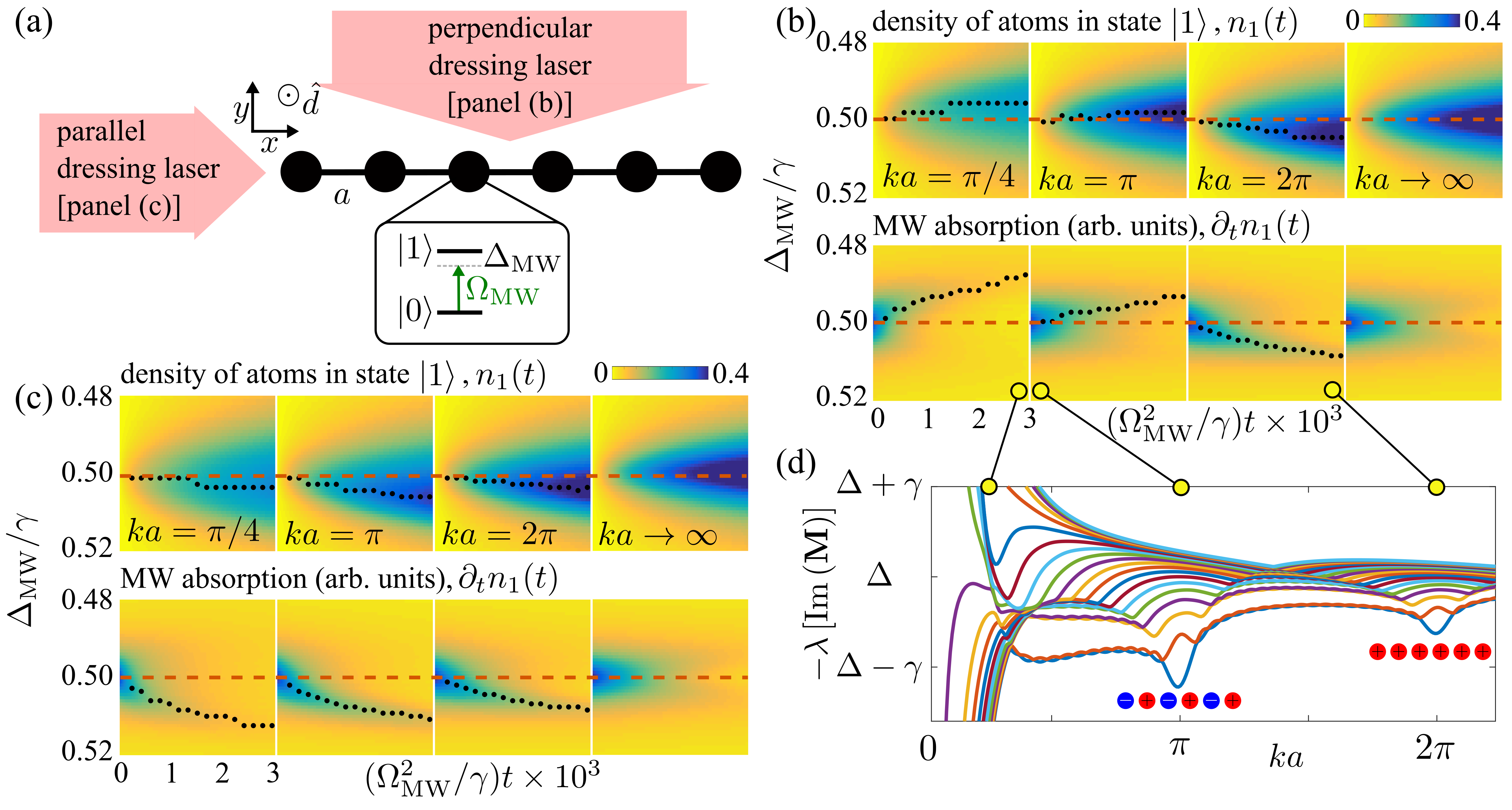}
  \caption{\textbf{MW spectroscopy.} (a) Sketch of a one-dimensional chain of atoms with spacing $a$ oriented along the $x$-axis. The atomic transition dipoles are pointing into the $z$-direction, and the laser is either irradiated perpendicular or parallel to the chain. (b) Time-evolution of the excitation density (upper row) and the MW absorption rate (lower row) for various values of the lattice spacing $ka$ and the MW detuning $\Delta_\mathrm{MW}$. The black dotted line indicates the position of the maximum (peak) value at a given time. The laser parameters are $\Delta=50\gamma$ and $\Omega=5\gamma$ with the laser being irradiated perpendicular to the chain. The number of atoms is $N=40$. (c) Same as in (b), but with the laser irradiated parallel to the chain. (d) Eigenvalues of the imaginary part of the matrix $\mathbf{M}$ [Eq. (\ref{eq:M_matrix})] for $N=20$. The colored dots near $ka=\pi$ and $ka=2\pi$ indicate the phase pattern of the atoms in the lowest energy eigenstate at $ka=\pi$ and $ka=2\pi$, i.e. alternating vs. uniform.}
\label{fig:3}
\end{figure*}
The phenomenology is similar to the dressing of Rydberg states. The difference is that the strength of the dressed state potential is proportional to  $\Omega^2/\Delta^2$ rather than $\Omega^4/\Delta^4$ \cite{bouchoule2002}. This is a consequence of the fact that (typically) the dominant interaction between Rydberg states is not exchange but rather a density-density interaction, so that two atoms have to be virtually excited to interact. A further difference is that the interaction energy of a dressed many-body state $\left|\mathcal{C}\right>$ cannot be constructed as the sum of binary interactions \cite{sevinccli2014}. This is a direct consequence of the fact that the dressing laser (virtually) excites collective states whose structure strongly depends on the particular arrangement of atoms in state $\left|1\right>$. In Fig. \ref{fig:2} we display the three-body potential $\varepsilon_3=-\mathrm{Re}\left[\Delta_{\left|110\right>\left|111\right>}\right]-2 \varepsilon_2-\varepsilon_1$. It saturates when $kr\rightarrow 0$ while for large separations it behaves approximately as $\varepsilon_3\rightarrow \frac{\Omega^2}{2\Delta^3} (G^2_{12}-12V^2_{12})$.

\textit{Many-body MW spectroscopy ---} The dressed states can be probed by coupling the transition $\left|0\right>\rightarrow\left|1\right>$ with a MW field of detuning $\Delta_\mathrm{MW}$ and Rabi frequency $\Omega_\mathrm{MW}$ \cite{jau2016,volchkov2018}. This allows to probe the energy spectrum of the $\left|1\right>$ manifold, in contrast to optical spectroscopy of the $\left|1\right>\rightarrow\left|2\right>$ transition. We assume that the MW Rabi frequency is weak compared to the (collective) dephasing rate which allows us to adopt a description of the dynamics in term of rate equations \cite{cai2013,lesanovsky2013,everest2016}. The transition rate between two configurations $\left|\mathcal{C}\right>$ and $\left|\mathcal{C}'\right>$ is only non-zero when they differ by one atom in the $\left|1\right>$ state (e.g. the ones displayed in Fig. \ref{fig:1}b) and is given by
\begin{eqnarray*}
  \Gamma_{\mathcal{C}\rightarrow\mathcal{C}'}=\frac{2\,\Omega^2_\mathrm{MW} \mathrm{Im}\left(\Delta_{\mathcal{C}\mathcal{C}'}\right)}{\left[\mathrm{Im}\left(\Delta_{\mathcal{C}\mathcal{C}'}\right)\right]^2+\left[\Delta_\mathrm{MW}\,(n_{\mathcal{C}'}-n_{\mathcal{C}})-\mathrm{Re}\left(\Delta_{\mathcal{C}\mathcal{C}'}\right)\right]^2}.
\end{eqnarray*}
Here $n_\mathcal{C}$ is the number of atoms in state $\left|1\right>$ contained in configuration $\left|\mathcal{C}\right>$. The resulting rate equations permit the calculation of the density of atoms in state $\left|1\right>$, $n_1(t)$, as well as the MW absorption rate, with the latter being proportional $\partial_t n_1(t)$. In Fig. \ref{fig:3} we show the corresponding data for a chain (spacing $a$) of $40$ atoms. Here we consider two situations: one where the dressing laser is irradiated from the top, i.e. the laser phases are identical for each atom (Fig. \ref{fig:3}b), and one where the laser is irradiated from the left, i.e. the laser phase changes from atom to atom (Fig. \ref{fig:3}c). In both cases we consider four different interparticle separations, interpolating between the strongly interacting and the non-interacting limit ($ka\rightarrow\infty$).

For short excitation times, both $n_1(t)$ and the MW absorption signal, exhibit a peak near the MW detuning $\Delta^0_\mathrm{MW}=\Omega^2/\Delta=0.5\times \gamma$ (the light-shift in the non-interacting limit). As time passes the position of the peak (marked by a black dotted line in the individual sub-panels) departs from $\Delta^0_\mathrm{MW}$, indicating the presence of interactions. In the density plots of the excitation density one observes only a small shift away from $\Delta^0_\mathrm{MW}$ as time progresses. The most significant difference between the non-interacting and strongly interacting regime ($ka=\pi/4$) is that in the latter the stationary state value of the excitation density --- which in all cases is $0.5$ --- is reached much more slowly. However, interaction effects are far more pronounced in the MW absorption spectrum. We here observe a clear shift of the absorption line when interactions are present.

The direction of the shift depends on the angle between the dressing laser and the atomic chain. This effect can be qualitatively understood by inspecting the eigenvalues of the imaginary part of the matrix $\mathbf{M}$, Eq. (\ref{eq:M_matrix}), which lends itself to being interpreted as the energy of collective states that are off-resonantly excited by the dressing laser. In Fig. \ref{fig:3}d we show for the purpose of illustration these eigenvalues for a chain of $20$ atoms as a function of the lattice spacing $ka$, assuming that all of the $20$ atoms are in the state $\left|1\right>$ and thus participate in the dressing. Note, however, that the MW transfers population between the states $\left|0\right>$ and $\left|1\right>$ and therefore the number of atoms participating in the dressing varies in time, and so do that the precise shape of the excitation spectrum and conversely the energy shift due to the laser dressing.

A special situation is encountered when the lattice spacing is a multiple of $\pi/k$. Here, the excitation spectrum displayed in Fig. \ref{fig:3}d possesses pronounced dips with lower energy. They decrease effectively the detuning and thus enhance the coupling of the dressing laser to the collective states. These dips occur because at these points the interaction matrix $V_{\alpha\beta}$ [Eq. (\ref{eq:V_matrix})] has either only positive entries or the sign of the entries alternates as $(-1)^{\alpha-\beta}$. The eigenstates corresponding to the collective excitation with lowest energy thus have an alternating or uniform phase pattern, respectively, as depicted in Fig. \ref{fig:3}d through the red/blue circles. This phase pattern influences the coupling strength of the dressing laser to the collective states and thereby the collective light shift. This is particularly visible in the data shown in Fig. \ref{fig:3}b,c for $ka=\pi$. When the dressing laser is irradiated from the side, its phase pattern is alternating and therefore the time-dependent shift of the MW absorption is drastically different compared to the case in which the laser is perpendicular to the chain and each atom experiences a uniform laser phase. This is a clear signature for there being indeed collective excitations at the heart of the dressing of dense atomic gases. In contrast, such sensitivity to the laser phase does not occur in conventional Rydberg dressing as excited atoms interact here via a density-density interaction which is phase insensitive.

\textit{Conclusions and outlook ---} We have developed a theory of dressed dense atomic gases. We showed that generically these systems feature many-body interactions as well as delocalized excitations and discussed how interactions manifest themselves in MW spectroscopy and the evolution of the excitation density. In our numerical simulations we so far focussed on small ensembles in one dimension. The theory is, however, also applicable for higher dimensions. In the future it would be interesting to extend our study to large three-dimensional examples and investigate the potential of the dressed dense atomic gases to study coherent collective quantum dynamics with many-body interactions, moving away from the limit described via rate equations.

\begin{acknowledgments}
The research leading to these results has received funding from the European Research Council under the European Union's Seventh Framework Programme (FP/2007-2013)  [ERC Grant Agreement No. 335266 (ESCQUMA)] and from the European Union's H2020 research and innovation programme [Grant Agreement No. 800942 (ErBeStA)]. Funding was also received from the EPSRC [Grant No. EP/M014266/1]. IL gratefully acknowledges funding through the Royal Society Wolfson Research Merit Award. BO was supported by the Royal Society and EPSRC [Grant No. DH130145].
\end{acknowledgments}

%

\end{document}